\documentclass[aip,pop,reprint]{revtex4-1}

\usepackage{amsmath,amssymb,amsthm}
\usepackage{graphicx}

\newcommand{\mvec}[1]{\boldsymbol{#1}}
\newcommand{\ddx}[2]{\frac{{\rm d}#1}{{\rm d}#2}}
\newcommand{\ppx}[2]{\frac{\partial #1}{\partial #2}}
\newcommand{\grad}{\mvec{\nabla}}

\newcommand{\kperp}[1]{\bar{k}_{\perp #1}}
\newcommand{\kpar}{\bar{k}_{\|}}
\newcommand{\eavg}[1]{\left\langle #1 \right\rangle}

\begin{document}


\title{Weak Turbulence in the Magnetosphere:  Formation of Whistler Wave Cavity by Nonlinear Scattering}



\author{C. Crabtree}
\email[]{chris.crabtree@nrl.navy.mil}
\affiliation{Plasma Physics Division, Naval Research Laboratory,
  Washington, DC 20375-5346, USA}

\author{L. Rudakov}
\affiliation{Icarus Research Inc., P.O. Box 30870, Bethesda, MD
  20824-0780, USA}

\author{G. Ganguli}
\author{M. Mithaiwala}
\affiliation{Plasma Physics Division, Naval Research Laboratory,
  Washington, DC 20375-5346, USA}

\author{V. Galinsky} 
\author{V. Shevchenko}
\affiliation{University of California--San Diego, San Diego, CA, USA}

\date{}

\begin{abstract}
  We consider the weak turbulence of whistler waves in the in
  low-$\beta$ inner magnetosphere of the Earth.  Whistler waves with
  frequencies, originating in the ionosphere, propagate radially
  outward and can trigger nonlinear induced scattering by thermal
  electrons provided the wave energy density is large enough.
  Nonlinear scattering can substantially change the direction of the
  wave vector of whistler waves and hence the direction of energy flux
  with only a small change in the frequency.  A portion of whistler
  waves return to the ionosphere with a smaller perpendicular wave
  vector resulting in diminished linear damping and enhanced ability
  to pitch-angle scatter trapped electrons.  In addition, a portion of
  the scattered wave packets can be reflected near the ionosphere back
  into the magnetosphere.  Through multiple nonlinear scatterings and
  ionospheric reflections a long-lived wave cavity containing
  turbulent whistler waves can be formed with the appropriate properties to
  efficiently pitch-angle scatter trapped electrons.  The primary
  consequence on the Earth’s radiation belts is to reduce the lifetime
  of the trapped electron population.
\end{abstract}


\maketitle 

\section{Introduction}
\label{sec:intro}


The dynamics of whistler wave packets in the inner magnetosphere have
been investigated using linear ray tracing for decades
\cite{haselgrove55,Kimura66,lauben01,santolik01,bortnik03}.  These
studies are important in understanding the lifetime of trapped
energetic electrons in the Earth's radiation belts because pitch-angle
scattering by electromagnetic whistler waves into the loss-cone can be
the dominant loss mechanism\cite{abel98}.  The rate of pitch-angle
scattering is a sensitive function of the spatial, spectral, and
wave-normal distributions of the wave energy density, and therefore
mechanisms that control the distribution of wave energy are important.
Recently the results of \citet{hasagawa75} and \citet{tripathi77} have
been extended to demonstrate that, through nonlinear (NL) induced
scattering by thermal electrons in low-$\beta$ plasmas,
quasi-electrostatic lower-hybrid waves can scatter into
electromagnetic whistler waves\cite{ganguli10,mith11}.  In this paper
we report on initial efforts to incorporate this NL mechanism into the
calculation of the propagation of whistler waves in the plasmasphere.
We find that when the turbulent wave energy density exceeds a
threshold, NL scattering can scatter waves that have exhausted their
usefulness for pitch-angle scattering back into useful waves capable
of further pitch angle scattering.  We suggest a weak turbulence
framework for incorporating these effects into ray tracing
calculations.  This framework is a natural extension of linear
techniques such as ray tracing and quasi-linear diffusion that have
been the workhorse of radiation belt physics for decades.

Magnetospherically reflecting whistlers (MRW) are whistler waves,
possibly originating from lightning strikes, whose ray paths in the
magnetosphere roughly follow magnetic field lines, reflecting many
times between the magnetic poles.  Waves originating in the ionosphere
propagate to higher and higher altitude and the perpendicular
component of the wave vector increases\cite{Edgar76}.  Based on linear
theory there are two important physical effects that limit the degree
of influence that MRW are predicted to have on the lifetime of trapped
energetic electrons.  First, linear damping of whistler waves depends
strongly on the perpendicular wavelength (increasing as $k_\perp^2$).
Whistler waves have a significant parallel electric field when the
perpendicular wavelength becomes comparable with the electron skin
depth.  This parallel electric field is the driving force of linear
Landau damping (resonant interactions with electrons) as well as
collisional damping.  Also, electron drift $\mvec{E}\times\mvec{B}$
energy increases as the square of $k_\perp$.  Thus as whistler waves
propagate from their source region and the perpendicular component of
the wave vector becomes larger the linear damping of the waves becomes
stronger.  Second, the rate of pitch-angle scattering of trapped
electrons depends strongly on the perpendicular wavelength.  As the
perpendicular wavelengths become shorter than the electron skin depth
the whistler wave becomes more electrostatic in nature (more
lower-hybrid like) and thus the magnetic component (which pitch-angle
scatters through the Lorentz force, $(e/c)\mvec{v}\times\mvec{B}$) is
reduced.  In this article we will demonstrate that when the energy
density of whistler waves is large enough, NL induced scattering can
return a portion of short-wavelength quasi-electrostatic lower-hybrid
waves back into long-wavelength electromagnetic whistler waves which
are less damped and more effective at pitch-angle scattering.

The scattered waves (with smaller $k_\perp$) have only a slightly
lower frequency than the original whistler waves, but the direction of
$\mvec{k}$ and their trajectory through the magnetosphere can be
dramatically altered \cite{ganguli10,mith11}.  A portion of the
scattered waves can be returned back to the ionosphere, and some of
these waves will retrace the original ray paths back to the
magnetosphere.  These scattered waves are then able to efficiently
pitch-angle scatter relativistic electrons once again.  Through
multiple NL scatterings in the magnetosphere and reflections in the
ionosphere a cavity containing turbulent whistler waves may be formed
that is more long-lived than would be predicted without NL scattering,
because the effective $k_\perp$ and thus linear damping rate is
reduced.

In section \ref{sec:raytracing} we review the dynamics of whistler
wave packets without NL scattering from the ionosphere into the
magnetosphere and describe our assumptions in characterizing these
trajectories.  In section \ref{sec:theory}, we describe how we include
the effects of NL induced scattering into the dynamics of whistler
wave packets.  In section \ref{sec:cavity} we describe how a
long-lived wave energy cavity may be formed due to induced NL
scattering and describe some of the basic properties of this cavity.
In section \ref{sec:conclusions} we discuss the role that this newly
postulated wave cavity may play in radiation belt physics and outline
future efforts to improve our understanding of the properties of the
wave cavity.

\section{Ray Tracing Whistler Waves}
\label{sec:raytracing}

\subsection{Model Assumptions}

\begin{figure}
\includegraphics[width=\columnwidth]{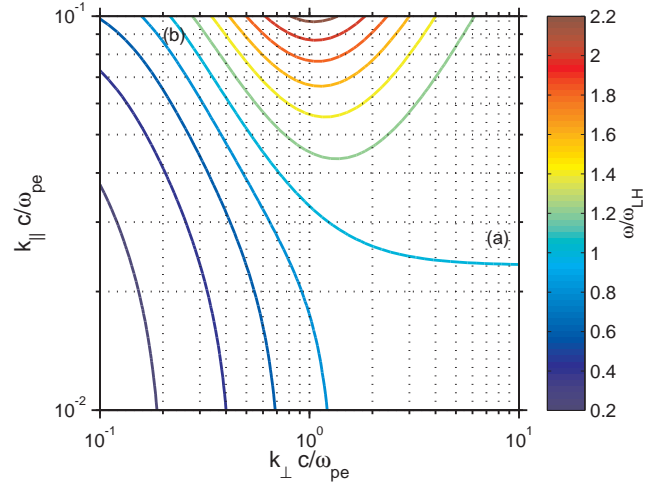}
\caption{\label{fig:disp} Contours of constant $\omega$ normalized to
  $\omega_{LH}$ calculated from Eq. \ref{eq:disp}}
\end{figure}

To model the dynamics of whistler wave packets in the
magnetosphere/ionosphere we use the following dispersion relation
applicable to cold plasmas for waves with frequencies between the
proton and electron cyclotron frequencies\cite{ganguli10},
\begin{equation}
  \label{eq:disp}
  \omega^2 =\left[
    \frac{\bar{k}_\|^2}{1+\bar{k}_\perp^2}\Omega_e^2+\omega_{LH}^2 \right]\frac{\bar{k}^2}{1+\bar{k}^2}
\end{equation}
where, $\omega_{LH}^2 = \sum_{ions} \omega_{ps}^2/(1 +
\omega_{pe}^2/\Omega_e^2)$ is the lower-hybrid frequency,
$\omega_{ps}$ is the plasma frequency, and $\Omega_s$ the cyclotron
frequency of species $s$.  In Fig. \ref{fig:disp}, we show contours
of constant frequency in
$(\kpar,\kperp{})=(k_\|,k_\perp)c/\omega_{pe}$ space calculated from
Eq. \ref{eq:disp}.  In the lower right corner ($\kperp{}>1$ and
$\kpar<\kperp{}(m_e/m_i)^{1/2}$) are the quasi-electrostatic
lower-hybrid waves with $\omega^2\simeq\omega_{LH}^2$, in the upper
left corner ($\kperp{}<1$ and $\bar{k}_\|>(m_e/m_i)^{1/2}$) are the
electromagnetic whistlers with
$\omega^2\simeq\kpar^2\bar{k}^2\Omega_e^2$, and in the lower left
corner ($\kperp{}<1$ and $\kpar<(m_e/m_i)^{1/2}$) are the magnetosonic
waves with $\omega^2\simeq\bar{k}^2\omega_{LH}^2\simeq k^2 V_A^2$,
where $V_A^2=B^2/(4\pi n_i m_i)$ is the Alfv\'{e}n velocity.

We use magnetic coordinates $(\chi,\alpha,\beta)$ where
$\mvec{B}=\grad\chi=\grad\alpha\times\grad\beta$ with a pure dipole
field such that,
\begin{equation}
\chi = -\frac{M\cos(\theta)}{r^2} \qquad
\alpha = \frac{M \sin^2(\theta)}{r} 
\qquad
\beta = \phi
\end{equation}
where $M=-8\times10^{25}$ G/cm$^3$ is the earth's dipole moment.  In
these coordinates the components of the group velocity may be
calculated,
\begin{equation}
  \begin{split}
    \dot{\chi} &= \ppx{\omega}{k_\chi} 
    =
    \frac{1}{1+\bar{k}^2}
    \left[\frac{\bar{k}^2}{1+\kperp{}^2}\Omega_e^2 +
      \frac{\omega^2}{\bar{k}^2}\right] \frac{c^2}{\omega_{pe}^2} \frac{g^{\chi\chi}k_{\chi}}{\omega}
    \\
    \dot{\alpha} &= \ppx{\omega}{k_\alpha}
    = \frac{1}{1+\bar{k}^2}\left[
    -\frac{\bar{k}_\|^2\bar{k}^2}{\left(1+\bar{k}_\perp^2\right)^2}
    \Omega_e^2
    +\frac{\omega^2}{\bar{k}^2} \right] \frac{c^2}{\omega_{pe}^2}
    \frac{g^{\alpha\alpha} k_\alpha}{\omega}
    \\
    \dot{\beta} &= \ppx{\omega}{k_\beta}
    = \frac{1}{1+\bar{k}^2}\left[
    -\frac{\bar{k}_\|^2\bar{k}^2}{\left(1+\bar{k}_\perp^2\right)^2}
    \Omega_e^2
    +\frac{\omega^2}{\bar{k}^2} \right] \frac{c^2}{\omega_{pe}^2}
    \frac{g^{\beta\beta} k_\beta}{\omega}
  \end{split}
\end{equation}  
where $g^{\chi\chi}$, $g^{\alpha\alpha}$, and $g^{\beta\beta}$ are
components of the metric tensor.

\begin{figure}
\includegraphics[width=\columnwidth]{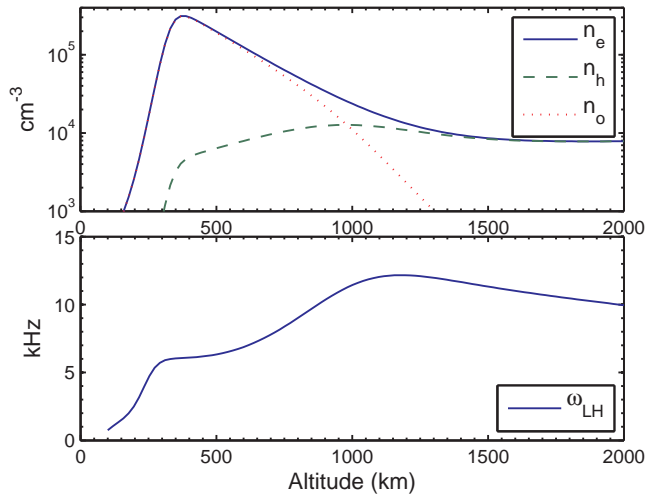}
\caption{\label{fig:model} Densities (top) and lower-hybrid frequency
  (bottom) vs altitude in model ionosphere.}
\end{figure}

To model the density in the magnetosphere we use a simple analytical
two ion species model that was designed to capture the main features
of a standard ionospheric density profile relevant to whistler
wave propagations.  The densities are all proportional to the magnetic
field, $ n_{s} = f_s(r) B$, multiplied by a function that depends
only on the radius.  The multiplying functions are defined by,
\begin{equation}
  \begin{split}
    f_e &= f_{e0} (r-r_0) + f_{e1}(1+\tanh(\frac{(r-r_1)}{\Delta r_1}))
    \exp(-\frac{r}{\Delta r_2}) 
    \\
    f_H &= \left[1/2(1+\tanh( \frac{r-r_{H1}}{\Delta r_{H1}}) \right] f_e
    \\
    f_O &= 1 - f_H
  \end{split}
\end{equation}
with the values of parameters determined by a fit to a night-time
ionosphere profile.  The densities and lower-hybrid frequencies are
plotted as a function of altitude in the upper ionosphere in Fig.
\ref{fig:model}.

In addition to following the dynamics of wave-packets through
$(\mvec{x},\mvec{k})$ space, we also follow the amount of energy lost
due to linear damping.  Since the region of interest (lower
magnetosphere) is a cold plasma ($T_e\sim 0.1-0.5 $ eV) Landau damping
by the thermal part is completely negligible compared to collisional
damping.  There sometimes exists a population of suprathermal
electrons that can be in resonance with whistler waves\cite{bell02},
however this damping is smaller than the collisional damping we
consider (see appendix \ref{sec:spth}).  We include the effects of
linear collisional (electron-ion) damping by using the approximate
damping rate,
\begin{equation}
  \label{eq:gl}
  \gamma_{L} = -\frac{\nu_{ei}}{2} \frac{2\kpar^2 + \kperp{}^2}{1+\kperp{}^2}
\end{equation}
where $\nu_{ei}=3\times10^{-5} n_e T_e^{-3/2}\,({\rm sec}^{-1})$, which requires a temperature for which we use a simple two temperature model profile given by,
\begin{equation}
  \label{eq:temp}
  T_e = 0.1+1/5(1+\tanh(\frac{\textrm{alt}-1100}{200}))\qquad\textrm{eV}
\end{equation}
where alt is the altitude in km.

For this initial work we confine ourselves to considering sources in
the ionosphere, however, we note that geomagnetic storms and chorus
emissions are examples of potential sources from the outer
magnetosphere.  As for sources in the ionosphere that could generate
whistler wave power there are many candidates, e.g. lightning,
VLF transmitters, and velocity ring distributions generated naturally
or man made such as in the Buaro experiment\cite{koons81} or an
envisioned future experiment \cite{ganguli07}.

\subsection{Linear Dynamics of Whistler Waves} 
\label{sec:lindynofwp}
  
\begin{figure}
  \includegraphics[width=\columnwidth]{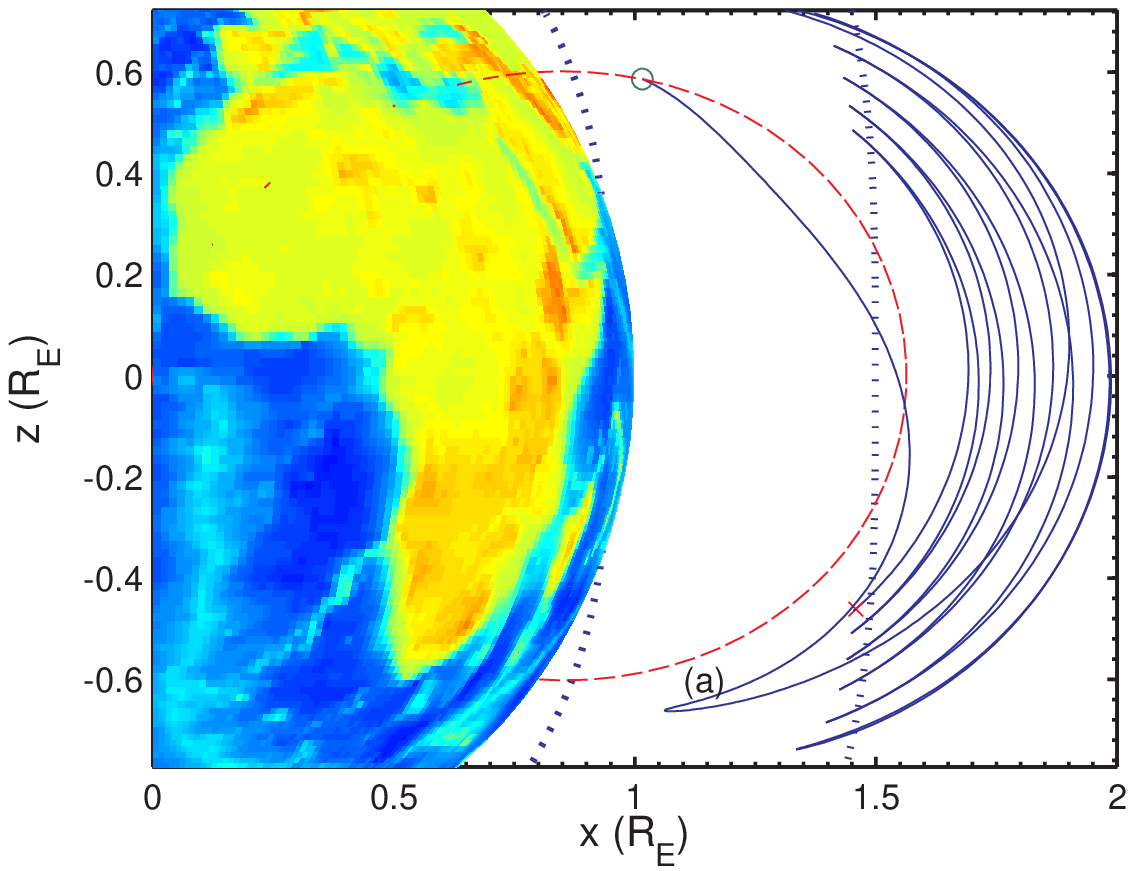}
  \includegraphics[width=\columnwidth]{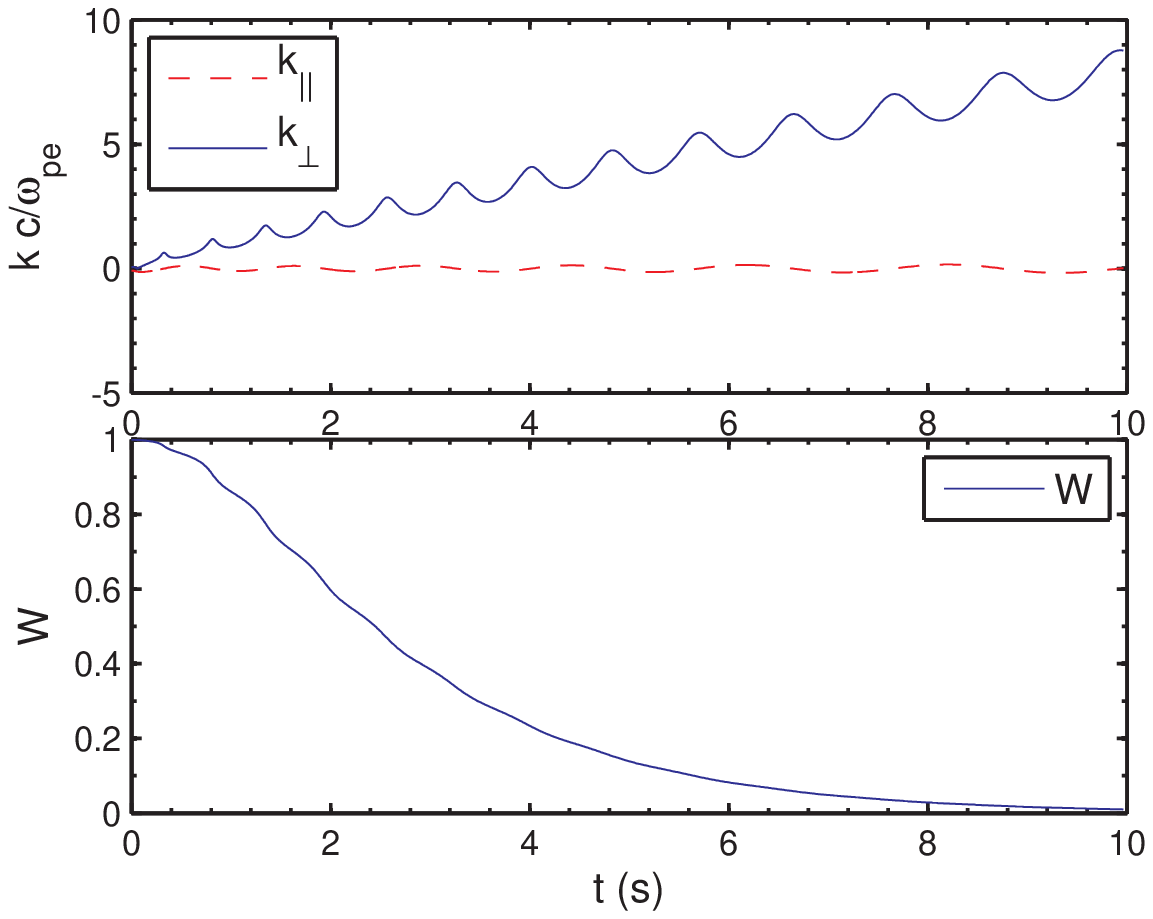}
  \caption{Wave packet trajectory in x-z plane for 6 kHz wave launched from 1100 km altiitude and 30$^{\circ}$N (top) and evolution of $\mvec{k}$ and energy (bottom).}
  \label{fig:NoScatter}
\end{figure}

\begin{figure}
  \includegraphics[width=\columnwidth]{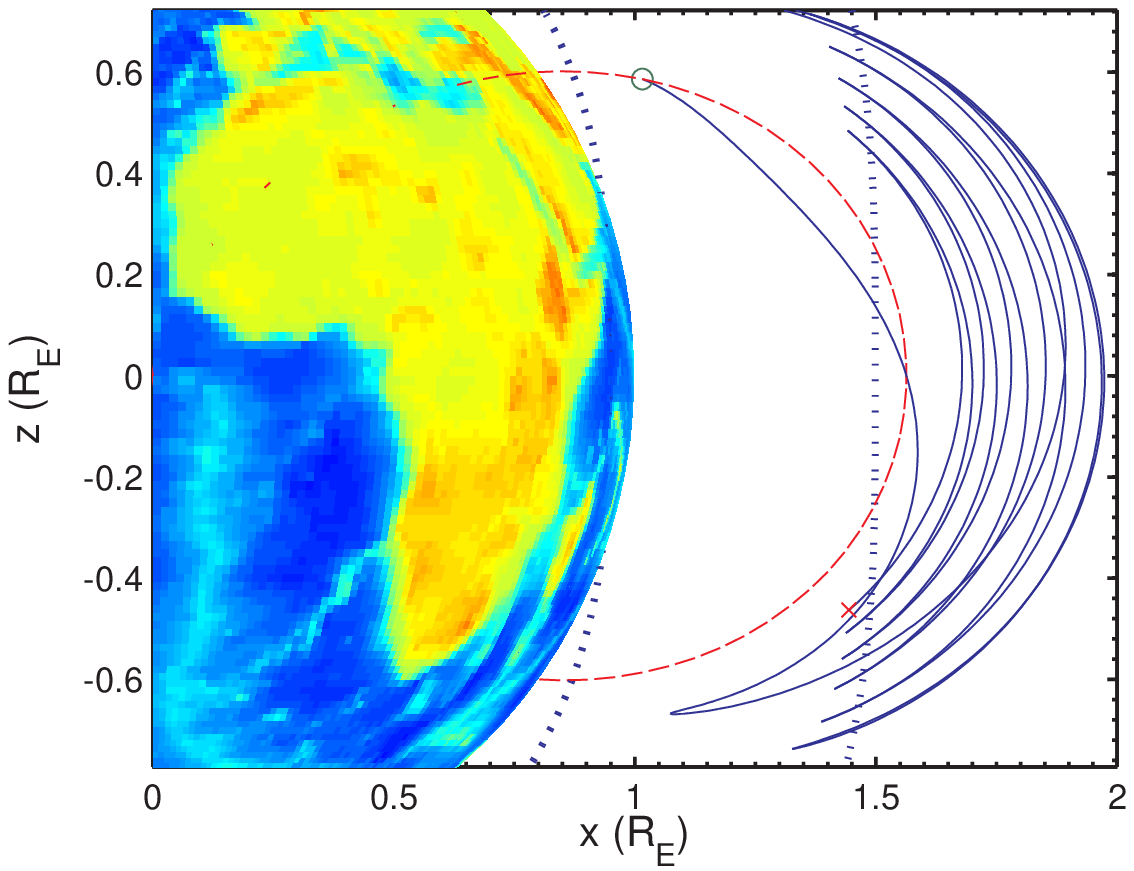}
  \includegraphics[width=\columnwidth]{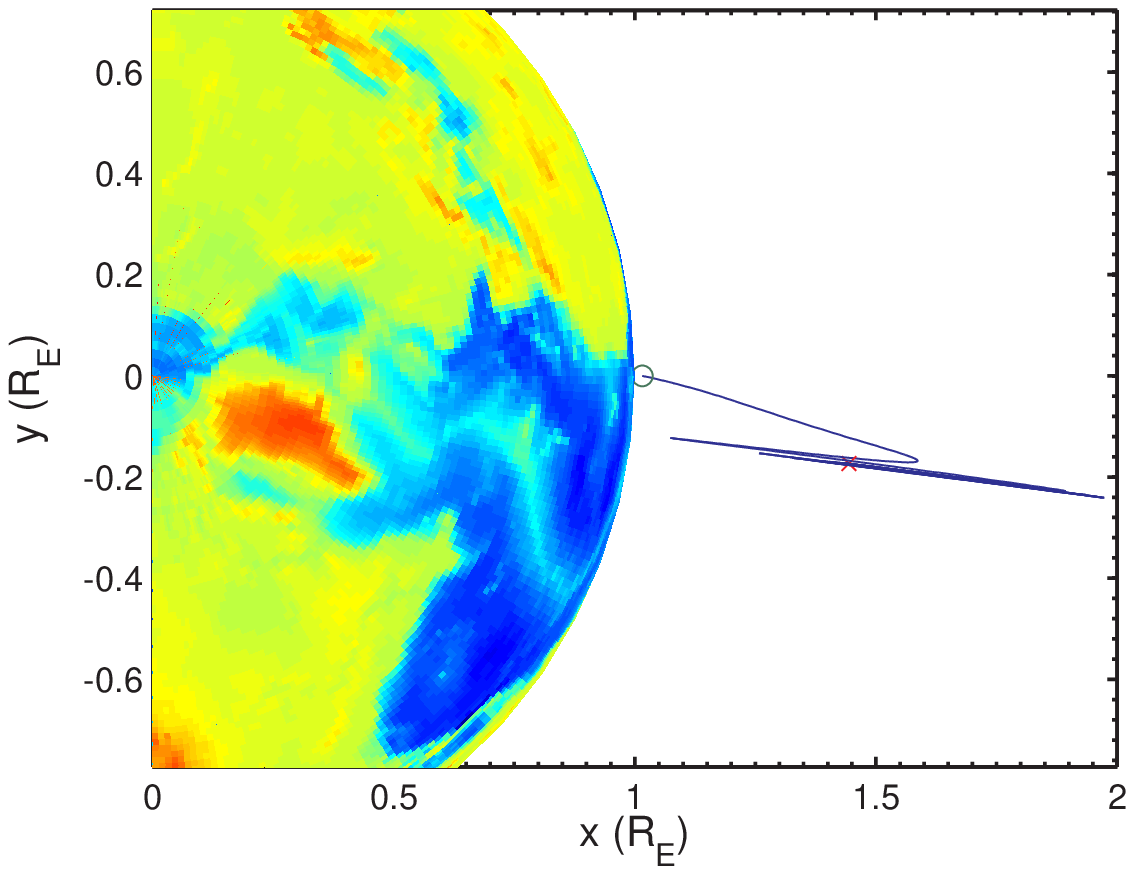}
  \caption{Wave packet trajectory for 6 kHz wave launched from 1100 km altitude and 30$^{\circ}$N with component of $\mvec{k}$ in azimuthal direction.  Top is x-z plane.  Bottom is x-y plane.}
  \label{fig:NoScatter3D}
\end{figure}

Generally speaking, whistler waves with small wave normal angles
propagate from their source region roughly along magnetic field lines,
reflecting between the magnetic poles and increasing the value of the
perpendicular wave vector.  The obliqueness of the wave allows for
cross-field propagation so that wave packets initiated in the
ionosphere propagate to higher and higher altitude (and L-shell).
Eventually the wave-packet reaches an L-shell greater than the L-shell
at which the frequency of the wave matches the lower-hybrid frequency
and the radial group velocity reverses.  By this time (a few seconds)
the wave packet has a large perpendicular wave vector and follows the
field line more closely. Its radial group velocity is relatively small
and the wave packet slowly makes its way back to the lower-hybrid
resonant surface.  Fig. \ref{fig:NoScatter} shows the dynamics of a
single wave packet, characteristic of packets that reach the
magnetosphere, launched from an ionospheric source at an altitude of
1100 km at 30$^{\circ}$ N and midnight local time with a frequency of 6
kHz (where the local lower hybrid frequency is about 12 kHz).  The
model ionosphere/magnetosphere is independent of the azimuthal
coordinate and thus the azimuthal component of the wave vector is a
constant of motion.  Therefore the wave-packet in figure
\ref{fig:NoScatter} which was launched with no azimuthal component of
the wave vector is confined to the meridian it began in.  In Figure
\ref{fig:NoScatter3D} we show the trajectory of a wave packet launched
from the same point in space and the same frequency, but with a
component of the wave vector out of the noon-midnight plane.  In this
case the packet initially has a significant group velocity in the
azimuthal direction, but the packet quickly settles down into a plane
of constant local time and does not proceed to encircle the earth.
These packets last almost 10 seconds before the initial energy is
reduced by 99\% by collisional damping.

\section{Theory}
\label{sec:theory}

In the framework of weak turbulence theory the dynamics of whistler
wave packets in the magnetosphere may be described by the wave-kinetic
equation which gives the evolution of the whistler wave packet ``number
density'' $N=W/\omega$ where $W$ is the energy density and $\omega$ is
the frequency,
\begin{equation}
  \label{eq:wke}
  \ppx{N}{t}
  + 
  \grad_r\cdot\left(\ddx{\mvec{r}}{t} N\right) 
  +
  \grad_k\cdot\left(\ddx{\mvec{k}}{t} N\right)
  =
  \gamma_{NL} N
  +
  2\gamma_{L} N
  + 
  Q
\end{equation}
where the left side of (\ref{eq:wke}) expresses the conservation of
phase volume in $(\mvec{r},\mvec{k})$ space of wave packets (Liouville
theorem).  The evolution of the wave packets through phase space
$(\mvec{x},\mvec{k})$ are given by the usual ray tracing equations,
\begin{equation}
  \ddx{\mvec{r}}{t} = \ppx{\omega}{\mvec{k}}
  \; \rm{and}\;
  \ddx{\mvec{k}}{t} = -\ppx{\omega}{\mvec{r}}.
\end{equation}
On the right hand side of Eq. (\ref{eq:wke}) $Q$ is a source of wave
energy \textit{e.g.} lightning discharges, chorus waves propagating
into the plasmasphere, VLF transmitters, etc., $\gamma_{L}$ is the
linear damping/growth rates which along with the usual quasi-linear
equations for the distribution of particles would give the complete
self-consistent linear dynamics of waves and particles.  The new
physics is contained in $\gamma_{NL}$, the NL induced
scattering rate\cite{ganguli10,mith11},
\begin{multline}
  \label{eq:gnl}
  \gamma_{NL} = \frac{1}{\omega_{k2}} \frac{\bar{k}_2^2}{1+\bar{k}_2^2}
  \sum_{k_1} \frac{|E_{k1}|^2}{4\pi n m_e} 
  \frac{\left| \mvec{k}_1 \times \mvec{k}_2\right|^2_{\|}}{k_{\perp 1}^2 k_{\perp 2}^2}
  \\
  \times
  \frac{(\mvec{k}_2 - \mvec{k}_1)^2 \bar{k}_1^2}{1+\bar{k}_1^2} 
  \frac{{\rm Im} \epsilon^{e}_{k_1-k_2} |\epsilon^i_{k_1-k_2}|^2}{
    \left|\epsilon^e_{k_1-k_2} + \epsilon^i_{k_1-k_2}\right|^2}
\end{multline}
where $\epsilon^{e,i}_{k_1-k_2}$ are the plasma electron and ion
susceptibilities, and $E_{k1}$ is related to $N_{k1}$ (detailed
discussion is contained in the appendix).  $\gamma_{NL}$ is to be
interpreted as the rate of change of the energy contained in waves
with wave vector $(\kpar,\kperp{})$ due to all other waves with
different frequencies and wave vectors with
$(\bar{k}_{\|1},\kperp{1})$.  This formula was derived for isothermal,
$T_e\simeq T_i$, low-$\beta$ magnetospheric plasmas from the drift
kinetic equation for electrons with a fluid response for ions.  It is
the electromagnetic generalization of the lower-hybrid induced
scattering rate found by \citet{hasagawa75} and used in tokamak
heating scenarios.  The scattering rate can be physically understood
as the landau damping of thermal electrons by the electric field
structure created by the beating of two waves.  It is important to
point out that the nature of this electric field structure is
inherently 3D as can be seen by the presence of the cross product of
two perpendicular wave vectors and the parallel components of the wave
vectors within the plasma dispersion function.  2D simulations with
the simulation plane perpendicular to the magnetic field, or 2D
simulations with the magnetic field along one of the simulation axes
will miss this fast scattering rate \cite{ganguli10} and will
generally see only a slower scalar nonlinearity.  Three-dimensional
simulations, or 2D simulations where the magnetic field is oblique to
the simulation plane are required.  For the purposes of this paper,
the importance of this three-dimensionality is that NL scattering
allows wave packets with components of the wave vector in a meridional
plane to scatter out of the plane as will be shown in the next
section.  Equation (\ref{eq:gnl}) was developed in the framework of
weak turbulence theory, which considers the NL coupling of linear
waves with stochastic uncorrelated phases, so that all waves are
solutions to the same linear dispersion relation, and hence the NL
induced scattering essentially redistributes energy in
$\mvec{k}$-space\cite{sagdeev69,davidson72}.

From Eq. (\ref{eq:gnl}) we see that the NL scattering rate is
largest when the change in frequency $\Delta\omega_{k1} \simeq
v_{te}|k_{\|1}-k_{\|}|$, and since the electron temperature in the
plasmasphere is of the order of an electron volt (which gives an
electron thermal speed of about 4$\times10^{7}$ cm/s) then the change
in frequency due to a scattering is much smaller than the change in
wave vector.  This allows one to view the lines of constant frequency
in Fig \ref{fig:disp} as all of the possible waves that a given wave
could scatter into.  For example, if we have a wave that is
lower-hybrid like, marked with point (a) in the figure, it may scatter
into a whistler wave, marked with point (b) in the figure.  This has
important consequences for the influence of whistler waves on the
radiation belts, namely that waves that have already become
lower-hybrid like due to the evolution of individual wave packets in a
dipole field as predicted by ray tracing, have an opportunity to
become whistler waves again.

To gain a detailed understanding of the effects of the NL
scattering on the propagation of wave power in the radiation belts a
self-consistent solution, either numerical or analytical, to the
wave-kinetic equation is needed.  Since this is beyond the scope of
this work, we adopt a test wave-packet approach to investigate the
effects of NL scattering and thus solutions to Eq. \ref{eq:wke}.

\section{Effect of NL Scattering on Wave Packet Trajectories}
\label{sec:cavity}

In this section we describe how a cavity in the magnetosphere may be
formed when the wave energy density exceeds a threshold, we describe
the properties of the cavity, and estimate the influence the formation
of the cavity will have on the trajectory and fate of whistler wave
packets and on the lifetime of trapped energetic electrons.

\subsection{Access to magnetospheric cavity from ionospheric source}

\begin{figure}
  \includegraphics[width=\columnwidth]{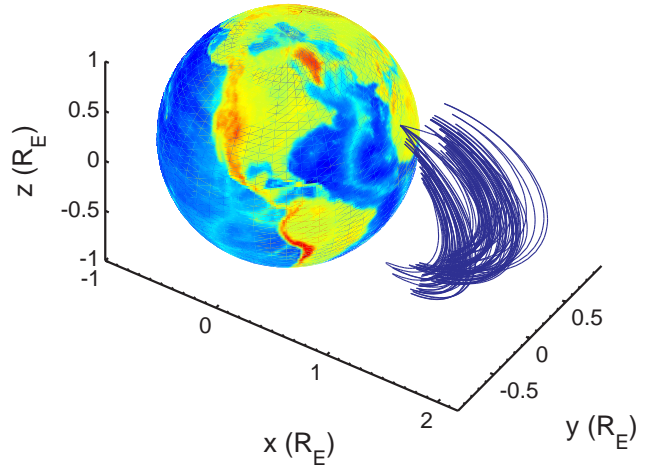}
  \caption{Trajectories of wave packets that enter magnetospheric
    cavity launched from 1100 km altitude, 30$^{\circ}$ N, with a
    frequency of 6 kHz.}
  \label{fig:MagEntry}
\end{figure}

\begin{figure}
  \includegraphics[width=\columnwidth]{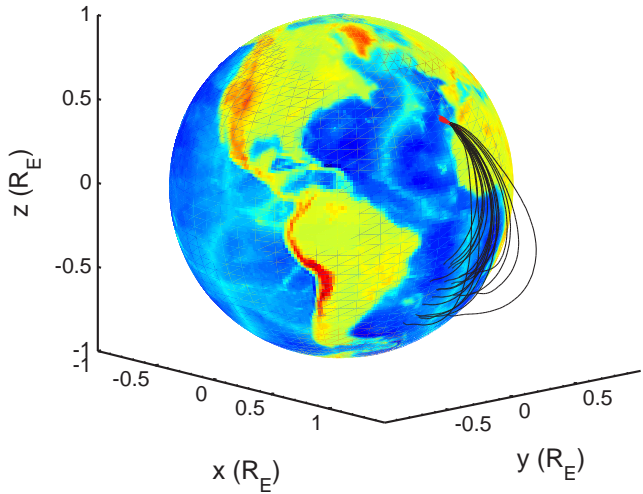}
  \caption{Trajectories of wave packets that are lost to the Earth
    launched from 500 km altitude, 30$^{\circ}$ N, with a frequency of
    6 kHz.  Red paths are lost to the Northern hemisphere.  Black
    paths are lost to the Southern hemisphere.}
  \label{fig:IonosphericEntry}
\end{figure}

We choose a source region in the ionosphere or in the transition
region between the ionosphere and the magnetosphere (see Table
\ref{tab:AccessCavity}) and suppose that the whistler energy flux is
nearly monochromatic and isotropic in $\mvec{k}$ space.  Launching
whistler wave packets with the same frequency, from the same location
in space, with random directions of the wave vector we can build up a
background.  By choosing a random direction for the wave vector many
packets will immediately be lost by propagation towards the Earth's
surface, and another percentage of packets will be lost by closely
following the magnetic field lines and reaching the Earth's surface in
the opposite hemisphere.  In addition, some of the packets are trapped
by the steep electron density gradient (closest to the peak in density
near 400 km), and because of the high density are quickly
collisionally damped.  We call these packets ducted.  The percentages
of packets for the different cases are listed in table
\ref{tab:AccessCavity}.  In Fig. \ref{fig:MagEntry} we show the
trajectories of wave packets that enter the magnetospheric cavity
launched from 1100 km altitude, 30$^{\circ}$ N with a frequency of 6
kHz.  This figure gives an idea of the three dimensional geometry of
the cavity which will be further discussed in the section
\ref{sec:propcav}.  In Fig. \ref{fig:IonosphericEntry} we show a
sample of wave packet trajectories launched from 500 km altitude with
a frequency of 6 kHz that are lost to the Earth.  We find that the
percentage of packets that enter the magnetospheric cavity increases
for the higher altitude and latitude initial conditions, mainly
because more packets that travel to the southern hemisphere are
refracted back into the magnetospheric cavity.
 
\subsection{Threshold for NL scattering}

\begin{figure}
  \includegraphics[width=\columnwidth]{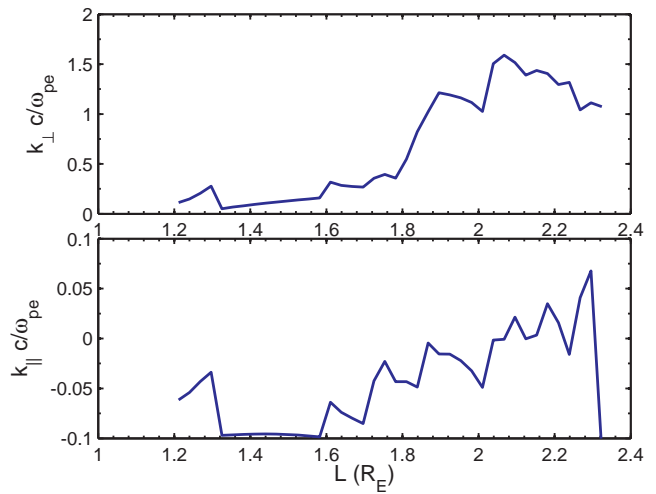}
  \caption{Ensemble average of the perpendicular wave vector for waves launched from 500 km altitude with a frequency of 4 kHz.}
  \label{fig:kperp}
\end{figure}

We estimate the threshold in wave energy density necessary for NL
induced scattering to play an important role in the evolution of
whistler waves.  To do this we suppose that the source lasts for a
time much longer than the bounce time of wave packets between the
magnetic poles and follow wave packets until they reach the surface of
the earth or their energy density reduces by 99\% while oscillating
near the lower hybrid surface in the magnetosphere.  From this
background of waves we can form a grid in space and calculate ensemble
averages of useful quantities such as $\kperp{}$.  We define the
following average,
\begin{equation}
  \label{eq:avg}
  \left\langle X \right\rangle(\mvec{x}_i) = 
  \frac{\sum_{\mvec{x}\in\mvec{x}_i} W X}{\sum_{\mvec{x}\in\mvec{x}_i} W }
\end{equation}
where $W$ is the energy of the particular wave-packet.  In Figure
\ref{fig:kperp} we plot such quantities from an ensemble of waves
launched from 500 km altitude with a frequency of 4 kHz (63\% of the
LH frequency).  Since we assume that the source lasts for 30 seconds we
record all equatorial plane crossings at different times to form this
average.  

Next, we estimate the NL scattering rate (Eq. \ref{eq:gnl}) by
assuming a broad band of turbulence, $\delta \omega \sim \omega$, and
replace the summation with integrations.  Using the condition that the
integrand is dominated by the condition $\zeta\sim 1$, and
the change in frequency is comparable to the ion-cyclotron frequency
(see Appendix \label{sec:app}) the NL scattering rate may be approximated as,
\begin{equation}
  \label{eq:agnl}
  \gamma_{NL} \sim \omega_{LH} \left(\frac{M}{m}\right)^{3/2}  \frac{\omega}{\delta \omega} 
  \frac{\eavg{\kperp{}}^8}{\left(1+\eavg{\kperp{}}^2\right)^2} \frac{B^2_{1}}{B_0^2}
\end{equation}
Then, from the form of Eq. (\ref{eq:wke}) we expect that NL scattering
will dominate linear damping when the energy density is large enough
such that $\gamma_{NL} > \gamma_{L}$.  Performing the same averaging
procedure on the collisional damping rate, Eq. (\ref{eq:gl}), we can
estimate the NL threshold as,
\begin{equation}
  \label{eq:nlthreshold}
  \left(\frac{B_1^2}{8\pi}\right)_{th} \sim \frac{1}{2} \frac{\nu_{ei}}{\omega_{LH}} \left(\frac{m}{M}\right)^{3/2} \left(\frac{1+\eavg{\kperp{}}^2}{\eavg{\kperp{}}^6}\right) 
\frac{\delta\omega}{\omega} \frac{B_0^2}{8\pi}
\end{equation}
in the limit $\eavg{\kpar}<\eavg{\kperp{}}$ which is justified by the
calculations leading to Fig. \ref{fig:kperp}.  Then taking the peak
$\eavg{\kperp{}}$ from the ray tracing calculations, and estimating
$\delta\omega/\omega\sim1/2$ we can then estimate the NL threshold
energy.  We find that the energy density to reach threshold
corresponds to magnetic field amplitudes of about 100 pT.  It should be
pointed out that this estimated threshold is of the same order of
magnitude as the observed amplitude of plasmaspheric hiss
\cite{thorne73,smith74}. Thus it can play an important role in such
phenomenon as radiation belt slot formation and it can effect theories
of the source of plasmaspheric hiss.  As we will show in the next
section NL scattering can significantly effect the wave propagation.

\subsection{Properties of Wave Energy Cavity}
\label{sec:propcav}
In the previous sections we computed the accessiblity of whistler
waves into the magnetosphere and the threshold in local wave energy
density necessary for NL scattering to be important.  Now we use ray
tracing calculations to estimate the volume of the cavity so that we
may estimate the energy necessary to be input to the cavity to achieve
sufficient NL scattering.  We launch 1000 wave packets into the
magnetosphere and compute the maximum extent in the azimuthal
direction (in units of hours, \textit{i.e.} 12 hours is 180 degrees),
and the maximum extent in latitude.  In table
\ref{tab:DimensionCavity} we summarize the results of our
calculations.  In general we find that the azimuthal extent of the
cavity decreases with increasing frequency and altitude, and is fairly
insensitive to latitude.  The meridional extent of the cavity
increases with the latitude and altitude of the whistler launch point
and frequency.  To compute the maximum radial extent we find the the
maximum radial value achieved by at least 10\% of the packets.  The
reason for this distinction is that for the higher latitude cases
there are a minority of packets ($<10\%$ that can achieve large radial
excursions on the first or second magnetospheric reflection).  We find
that in general the maximum radial extent decreases with frequency and
increases with latitude.  Also included in the table is an estimate of the volume of
the cavity where the minimum radius of the cavity is always taken to
be 1.4 $R_E$.  From these volume estimates we can then use the wave
energy density necessary to overcome the NL threshold and arrive at
the necessary total energy to achieve threshold.  This is indicated in
the final column of table \ref{tab:DimensionCavity}.

\begin{figure}
  \includegraphics[width=\columnwidth]{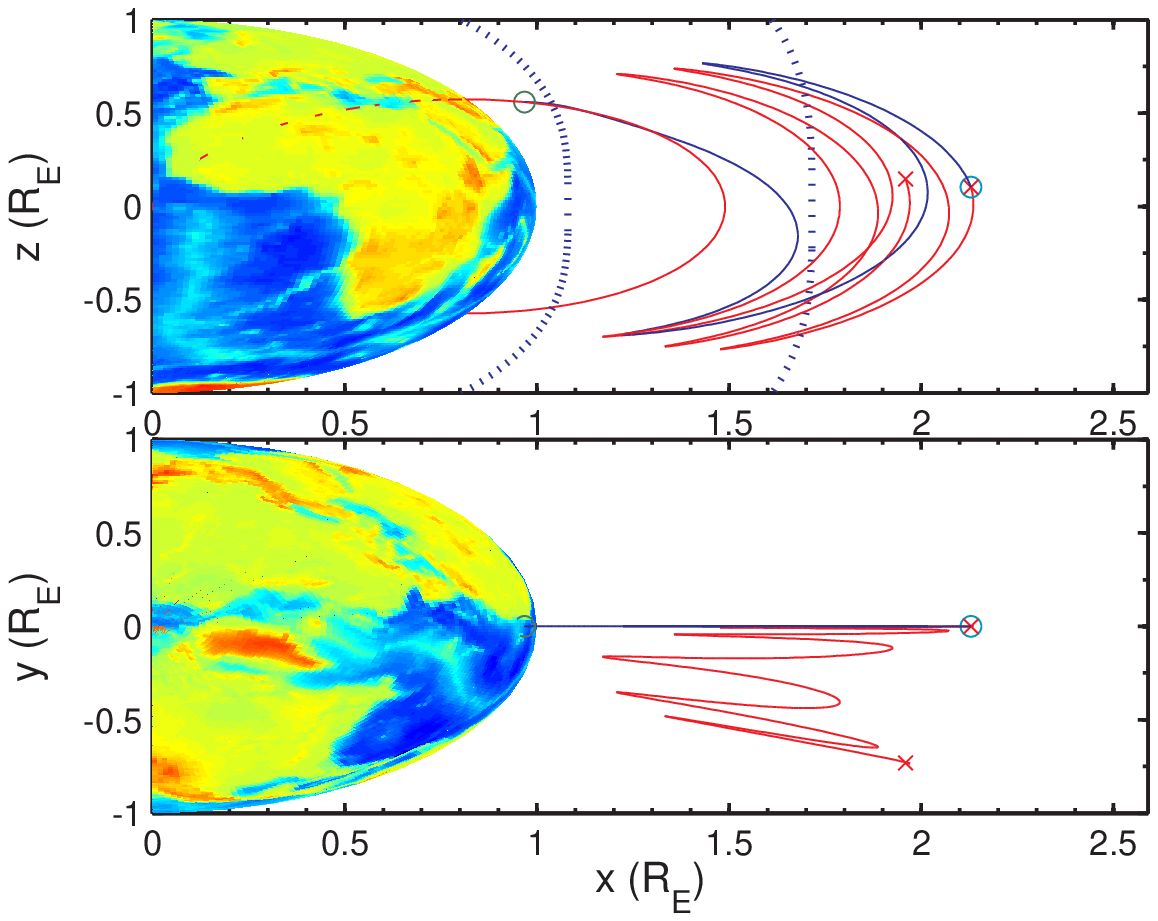}
  \includegraphics[width=\columnwidth]{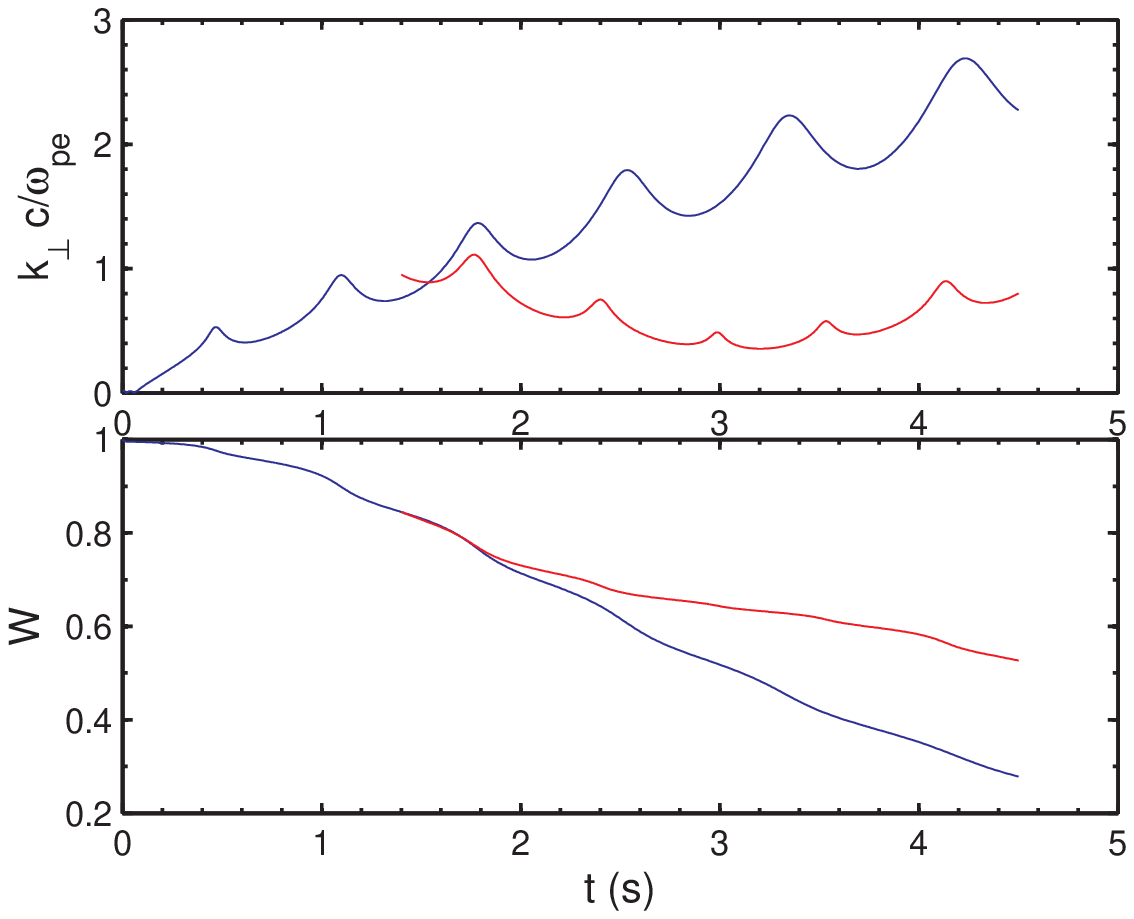}
  \caption{Wave packet trajectory (top) for 4 kHz wave launched from 750 km altiitude and 30$^{\circ}$N (blue) with wave vector confined to meridional plane and scattered after 1.4 seconds (red).  Perpendicular wave vector and energy of wave as a function of time (bottom), blue indicates unscattered wave and red is scattered wave.}
  \label{fig:WithScatter}
\end{figure}

Next we describe an example of a single scattering.  For this purpose
we assume that enough wave energy density is present in the system so
that a wave packet may scatter.  In Fig. \ref{fig:WithScatter} we
show the trajectory (top in blue) of a wave packet with a frequency of
4 kHz launched from 750 km altitude at 30$^{\circ}$ N latitude.  This
wave packet has the wave vector in the noon-midnight meridian so that
the wave packet is confined to the same meridian.  After 1.4 seconds
the wave packet has increased its value of $\bar{k}_{\perp}$ to about
unity and experiences a large angle scattering, such that the new wave
packet has a component of the wave vector out of the meridian plane.  The new
scattered packet also has $\bar{k}_{\perp}\sim1$, but this packet is
on a different trajectory so that $\bar{k}_{\perp}$ decreases in time
for a while before returning back to the magnetospheric lower hybrid
surface and increasing $\bar{k}_{\perp}$ again.  This reduction in
$\bar{k}_{\perp}$ allows for the energy density in the wave to be
almost a factor of 2 larger than the energy density found in the
packet without scattering (and a subsequent reduction in the final
value of $\bar{k}_{\perp}$).

\begin{center}
\begin{figure}
  \includegraphics[width=\columnwidth]{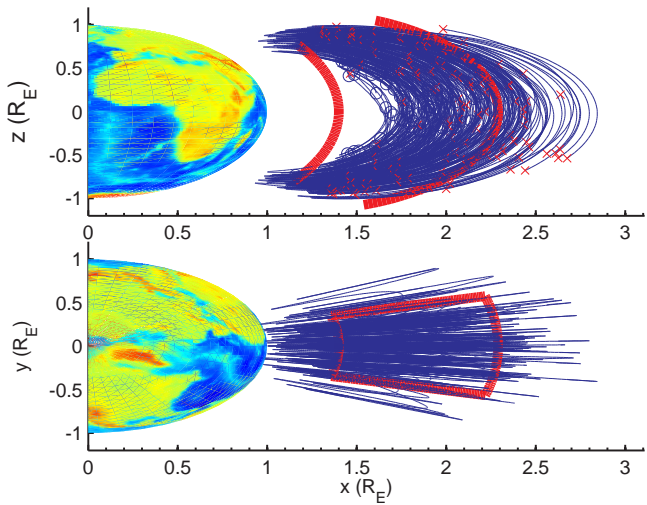}
  \caption{Scattered wave packet trajectories for 6 kHz wave launched
    from 1100 km altiitude and 30$^{\circ}$N. Solid thick lines (red)
    indicate approximate boundary of unscattered cavity.}
  \label{fig:albedo}
\end{figure}
\end{center}

Finally, we discuss the effects of many scatterings.  In figure
\ref{fig:WithScatter} we show what happens to a particular wave
packet that is scattered which says in the magnetospheric cavity.
However not all packets stay in the cavity.  Some are lost to the
Earth, and others experience large linear damping.  To begin to
quantify this energy loss we define a whislter wave albedo that we
calculate by launching whistler wave packets with random directions
for the wave vectors from inside the magnetospheric cavity and
integrating the ray tracing equations for a specified time and then
comparing the total energy of the packets at this time to the total
initial energy in the packets.  Thus, the albedo is defined as,
\begin{equation}
  \label{eq:albedo}
  Albedo = \frac{\sum W_{t=t_{end}}}{\sum W_{t=0}},
\end{equation}
where the numerator does not include packets that are lost to the
earth.  A good choice for the ending time in the albedo calculation
would be the time a particular packet spends before it's next
``collision'', \textit{i.e.} $1/\gamma_{NL}$.  For simplicity, we chose
the energy density such that the NL collision time was 1.5 seconds for
all packets.  In Fig. \ref{fig:albedo} we show a sample of ray
tracing calculations for which the wave packets remain in the cavity,
with their initial conditions marked by circles and their termination
points by x's.  Performing these kinds of calculations for all of the
cases we considered we summarize the results in table
\ref{tab:AlbedoCavity}.  In general we find that the albedo decreases
with increasing frequency, is fairly insensitive to the altitude of
release, and is larger for the higher latitude release case.  In
addition to the albedo the table shows the percentage of packets that
are lost to the Earth.  In all cases this is the dominant source of
energy loss.

Another feature of many NL scatterings is that the size of the cavity
can increase, this feature can be seen in the example figure
\ref{fig:albedo}, where the size of the new cavity can be seen by the
ray trajectories in blue overlaid on the original size of the cavity
estimated as thick red line in the figure.  One mechanism for
increasing the size is that packets which have become confined to a
particular meridian may be scattered such that they propagate
azimuthally.  An individual case can be seen in
Fig. \ref{fig:WithScatter}.  In table \ref{tab:AlbedoCavity} we
estimate the dimensions of the cavity after a first scattering.

\subsection{Effect of Cavity Formation on Energetic Trapped Electrons}

There are three main components to enhancing the pitch-angle
scattering of energetic trapped electrons by NL scattering of whistler
waves.  The first is that the energy density of the whistler waves
from a given source can be enhanced above the value predicted by ray
tracing without NL scattering.  This was demonstrated in figure
\ref{fig:WithScatter}, a corollary of this phenomenon is that the
waves can live longer in the cavity with NL scattering than without,
and this was discussed in relation to the albedo of the Earth.  The
final ingredient is that the average wave-normal angle can be
decreased, which enhances the pitch-angle scattering in the following
way.

The pitch-angle diffusion coefficient is approximately given by,
\begin{equation}
  \label{eq:dtt}
  D_{\theta\theta} \sim \frac{\Omega_e}{\gamma_R} \frac{W}{B_0^2/(8\pi)} 
  \left(\frac{J_1(\sigma)}{\sigma}\right)^2
\end{equation}
where $\sigma = k_\perp v_\perp/(\Omega_e/\gamma_R)$ and $\gamma_R$ is
the relativistic factor.  The cyclotron resonance condition
($\omega-k_{\|}v_{\|}=\Omega_e/\gamma_R$) for relativistic electrons
and whistler waves gives $k_{\|}c\cos(\theta) \simeq
\Omega_e/\gamma_R$ where $\theta$ is the electron pitch-angle.  With
this condition the argument of the Bessel function becomes $\sigma
\simeq (k_\perp/k_\|) \tan(\theta)$ which quickly becomes large due to
linear ray tracing of wave packets.  In this limit, $\sigma>>1$, the
pitch-angle diffusion coefficient becomes,
\begin{equation}
  \label{eq:htt2}
  D_{\theta\theta} \sim k_\| c \left(\frac{k_\|}{k_\perp}\right)^3 \frac{W}{(B_0^2/(8\pi)} \cos(\theta)\cot^3(\theta)
\end{equation}
which can be very small.  What this formula shows, is that without NL
scattering of whistlers the time with which a wave packet can
effectively scatter relativistic electrons is very short.  However,
with NL scattering a population of packets can be scattered back so
that the increase in $k_{\perp}$ from ray tracing is reversed, and the
rate of pitch-angle scattering can remain significant.

\begin{center}
\begin{table*}[ht]
  \hfill{}
  \begin{tabular}{|c|c|c|c|c|c|c|c|}
    \hline
    Altitude & Latitude & Frequency & LH Frequency 
    & \% into Cavity & \% into North & \% into South & \% Ducted 
    \\
    (km) & ($^{\circ}$ N) & (kHz) & (kHz) & (\%) &  (\%) & (\%) & (\%) 
    \\
    \hline
    \hline
    500 & 30 & 4 & 6.3 & 8.5 & 63 & 26 & 2 
    \\
    \hline
    500 & 30 & 6 & 6.3 & 8.0 & 57 & 35 & 0.4 
    \\
    \hline
    1100 & 30 & 4 & 12 & 41 & 52 & 7.6 & 0 
    \\
    \hline
    1100 & 30 & 6 & 12 & 38 & 52 & 10 & 0 
    \\
    \hline
    1100 & 30 & 8 & 12 & 38 & 52 & 10 & 0 
    \\
    \hline
    750 & 30 & 7 & 8.4 & 14 & 52 & 35 & 0 
    \\
    \hline
    850 & 30 & 7 & 9.7 & 18 & 51 & 31 & 0
    \\
    \hline
    950 & 30 & 7 & 10.9 & 25 & 52 & 23 & 0
    \\
    \hline 
    750 & 45 & 7 & 9.9 & 47 & 53 & 0 & 0
    \\
    \hline
    500 & 45 & 4 & 7.5 & 35 & 63 & 0 & 3
    \\
    \hline
    750 & 45 & 8 & 9.9 & 46 & 54 & 0 & 0
    \\
    \hline 
    600 & 45 & 6 & 8.2 & 48 & 52 & 0 & 0
    \\
    \hline
  \end{tabular}
  \hfill{}
  \caption{Access to magnetospheric cavity computed from 1,000 ray tracing calculations for each case.  }
  \label{tab:AccessCavity}
\end{table*}
\end{center}

\begin{center}
\begin{table*}[ht]
  \hfill{}
  \begin{tabular}{|c|c|c|c|c|c|c|c|}
    \hline
    Altitude & Latitude & Frequency & Azimuthal Extent & Maximum Radius & Latitudinal Extent & Volume
    & Energy 
    \\
    (km) & ($^{\circ}$) & (kHz) & (Hrs) & ($R_E$) & ($^{\circ}$) &($R_E^3$) & (MJ) 
    \\
    \hline
    \hline
    500 & 30 & 4 & 2.5 & 2.3 & 61 & 2.1 & 0.5
    \\
    \hline
    500 & 30 & 6 & 2.0 & 2.2 & 65 & 1.4 & 0.4
    \\
    \hline
    1100 & 30 & 4 & 2.3 & 2.4 & 64 & 2.4 & 0.6 
    \\
    \hline
    1100 & 30 & 6 & 1.9 & 2.2 & 68 & 1.6 & 0.4 
    \\
    \hline
    1100 & 30 & 8 & 1.7 & 2.2 & 71 & 1.3 & 0.3
    \\
    \hline
    750 & 30 & 7 & 1.9 & 2.2 & 68 & 1.3 & 0.3
    \\
    \hline
    850 & 30 & 7 & 1.9 & 2.2 & 69 & 1.5 & 0.4
    \\
    \hline
    950 & 30 & 7 & 1.8 & 2.2 & 69& 1.4 & 0.4
    \\
    \hline 
    750 & 45 & 7 & 2.2 & 3.1 & 89 & 7.4 & 1.9
    \\
    \hline
    500 & 45 & 4 & 2.6 & 3.3 & 88 & 10.8 & 2.7
    \\
    \hline
    750 & 45 & 8 & 2.2 & 3.0 & 89 & 6.9 & 1.7
    \\
    \hline 
    600 & 45 & 6 & 2.4 & 3.0 & 89 & 7.0 & 1.8
    \\
    \hline
  \end{tabular}
  \hfill{}
  \caption{Dimension of magnetospheric cavity computed from 1,000 ray tracing calculations for each case}.
  \label{tab:DimensionCavity}
\end{table*}
\end{center}

\begin{center}
\begin{table*}[ht]
  \hfill{}
  \begin{tabular}{|c|c|c|c|c|c|c|c|c|}
    \hline
    Altitude & Latitude & Frequency & Lost & Albedo & Azimuthal & Maximum & Latitudinal & Volume
    \\
    & & & & & Extent & Radius & Extent & 
    \\
    (km) & ($^{\circ}$) & (kHz) & \% & \%  & (Hrs) & ($R_E$) & ($^{\circ}$) &($R_E^3$) 
    \\
    \hline
    \hline
    500 & 30 & 4 & 22 & 65 & 4.6 & 2.8 & 93.5 & 10.6
    \\
    \hline
    500 & 30 & 6 & 47 & 38 & 3.5 & 2.5 & 88 & 5.4 
    \\
    \hline
    1100 & 30 & 4 & 19 & 68 & 4.3 & 2.9 & 99 & 12
    \\
    \hline
    1100 & 30 & 6 & 39 & 45 & 3.4 & 2.6 & 95 & 6.4
    \\
    \hline
    1100 & 30 & 8 & 57 & 28 & 2.6 & 2.4 & 88 & 3.5
    \\
    \hline
    750 & 30 & 7 & 51 & 34 & 3.1 & 2.5 & 87 & 4.4
    \\
    \hline
    850 & 30 & 7 & 50 & 34 & 3.1 & 2.5 & 89 & 4.5
    \\
    \hline
    950 & 30 & 7 & 50 & 34 & 3.4 & 2.5 & 89 & 5
    \\
    \hline 
    750 & 45 & 7 & 32 & 49 & 4 & 3.0 & 115 & 13.4
    \\
    \hline
    500 & 45 & 4 & 17 & 70 & 5.4 & 3.0 & 122 & 21 
    \\
    \hline
    750 & 45 & 8 & 37 & 43 & 3.2 & 2.9 & 110 & 9.4
    \\
    \hline 
    600 & 45 & 6 & 29 & 53 & 4 & 3 & 112 & 14
    \\
    \hline
  \end{tabular}
  \hfill{}
  \caption{Properties of magnetic cavity after the first scattering.}
  \label{tab:AlbedoCavity}
\end{table*}
\end{center}

\section{Conclusions}
\label{sec:conclusions}

In conclusion, we have shown that NL induced scattering can alter the
propagation of whistler wave energy in the magnetosphere from
ionospheric sources, such that a long-lived cavity may form that can
more effectively pitch-angle scatter trapped relativistic electrons
and affect the lifetime of the trapped population.  We have introduced
a framework for incorporating this NL effect into future more detailed
studies, and have used ray tracing calculations to explore the
properties of the NL cavity.  Next we highlight a couple of
observations from radiation belt physics for which NL scattering may
play an important role.

Plasmaspheric hiss, which plays an important role in pitch-angle
scattering, refers to observed broadband, electromagnetic, incoherent,
and turbulent fluctuations on the whistler branch mostly confined
to the high-density, cold plasmasphere. Typical amplitudes of the
magnetic fluctuations range from 10 to 100 pT\cite{thorne73,smith74},
\textit{i.e.} in the range where the NL threshold could be met. The
observation of the wave normal angles of hiss is the most contentious
and difficult to make.  Wave normal angles have been observed to be
(1) rather oblique, within 45 degrees of perpendicular to the magnetic
field\cite{sonwalkar88}, (2) parallel propagating\cite{santolik01} (in
both directions), and (3) a mixture of both types (1) and
(2)\cite{hayakawa86,storey91}.  Small wave-normal angles of
plasmaspheric hiss have been difficult to explain theoretically,
because ray tracing studies (as shown in Section \ref{sec:lindynofwp}) show that
$k_\perp$ becomes large very quickly.  The most established mechanism
involves refraction of whistlers off of the density gradient at the
plasmapause, however there are observations where the plasmapause may
be very distant.  The presence of these small-wave normal angle
whistlers may be due to NL scattering provided the energy density is
sufficiently high.  The source of plasmaspheric hiss has been
attributed to lightning\cite{sonwalkar89,draganov92}, chorus emissions
outside of the plasmasphere \cite{chum05,bortnik08}, and amplification
of background or embryonic waves by anisotropic electron
distributions\cite{thorne73,church83}, and the debate about which
source is more likely has hinged in some part to the azimuthal
accessibility and the wave-normal angle of waves all computed from
linear ray tracing.  As we have demonstrated both the wave-normal
angle and the azimuthal extent from a given source can be affected by
NL scattering.

As evidence of the importance of the wave-normal angle of
plasmaspheric whistler waves on the lifetime of energetic trapped
electrons \citet{meredith09} found that the lifetime of 2-6 MeV
electrons measured by SAMPEX in the inner slot region, that was filled
due to the Halloween storms of 2003, was barely influenced by
un-ducted whistler waves (with predicted large wave-normal angles),
however by including a small population of whistler waves with small
wave normal angles (assumed to be ducted whistlers) the observed
lifetimes of these electrons could be explained.

In the future, we plan to develop a more self-consistent solution to
the wave kinetic equation to investigate the evolution and equilibrium
of spectral wave-power in the radiation belts.  To compare these to
observational data more detailed models of the sources and equilibrium
structure of plasma density, and magnetic field would be useful.
Finally, we mention one future research topic which could have
profound consequences on the amplitude of turbulent whistler waves and
that is the inclusion of the growth of whistlers due to anisotropic
electron distributions (loss-cone distributions)\cite{sagdeev61} by
the mechanism put forward in application to magnetospheric physics by
\citet{kennel66}.  The criterion for growth by this mechanism has been
difficult to identify because, as was demonstrated in Section
\ref{sec:lindynofwp}, whistler waves quickly develop short
perpendicular wavelengths and experience large linear damping, and
thus waves would have only a short time to gain energy from the
unstable electron population.  However, with NL scattering whistlers
can spend a longer time with properties desirable for growth and thus
may better be able to tap into the unstable electron population.
Detailed analysis of this possibility will be done in the future.


%
%

%

\appendix

\section{Landau and Transit-Time Damping due to Suprathermal Electrons}
\label{sec:spth}


If Landau damping due to the suprathermal electrons (100 eV to about
1.5 KeV) is sufficiently large, then the wave energy required to
reach the threshold for NL scattering can potentially be very large.
In this appendix we calculate the collisionless (Landau and transit-time) damping
caused by the population of suprathermal electrons and show that it is
smaller than the collisional damping we have considered.
Consequently, for the parameters of interest the suprathermal
electrons will not play any significant role in the formation or
lifetime of the cavity within the plasmasphere.

Suprathermal electrons are observed as a sporadic
phenomenon\cite{bell02}.  To evaluate the damping rates we use
properties of the observed statistical average population of
suprathermal electrons.  Most studies focus on data from higher-L
shells than what are examined in our work (most of the ray-tracing
calculations in our study are confined to L$<$2.5).  For example,
Thorne and Horne\cite{thorne94} model the suprathermal population with
a series of Maxwellians between 4.2$<$L$<$5.7.  Bell \textit{et al.}
\cite{bell02} model the suprathermal electron population between
2.3$<$L$<$5.7 with power-laws and find that the density of
suprathermal electrons is about 200 times less (at 150 eV) than at the
higher L-shells.  Recently Li \textit{et al.}\cite{Li2010} reported on
global distributions of suprathermal electrons and confirmed that for
energies less than a few KeV the results of Bell et al. are more
appropriate for inside the plasmapause and that the density increases
with increasing L-shell.  Since our studies are at L-shells below
these observational works, it is likely that our calculations of
collisionless damping over-estimate the damping rates.  Finally, it
should be pointed out that if there is a population of suprathermal
electrons contributing to Landau damping, then consequently the
lifetime of these electrons in the radiation belts would be short as
these particles would be energized into the loss cone and precipitated
into the ionosphere.  Because of the sporadic nature of suprathermal
flux there will be periods with no resonant electrons to cause
collisionless damping.  However, to be conservative in the following
we assume that the population is always present and contributing to
collisionless damping

We calculate the damping rate by first calculating the wave power
dissipated by the energetic but tenous population of suprathermal
electrons,
\begin{equation}
  \label{eq:apb-1}
  P = \mvec{E}\cdot\mvec{J}=\frac{\omega}{8\pi}\mvec{E}^*\cdot\mvec{\chi}_a\cdot\mvec{E}
\end{equation}
where $\mvec{\chi}_a$ is the anti-hermitian part of the susceptibility
due to the suprathermal electrons.  By considering only Landau damping
($n=0$), taking $\gamma<\omega$ and $k_\perp \rho_e<<1$ then the limit
of the susceptibility tensor given by Stix\cite{stix} can be used to
calculate the power dissipated as
\begin{multline}
  \label{eq:apb-2}
  P = -\frac{\omega_{pe}^2}{8 n\omega} \int_0^{\infty} 2\pi v_\perp {\rm d}v_\perp
  \int_{-\infty}^{\infty} {\rm d}v_{\|} \delta\left(v_{\|} - \frac{\omega}{k_{\|}}\right)\ppx{f_0}{v_\|}
  \\
    \left[ 
      \frac{1}{4} \frac{k_\perp^2 v_\perp^4}{\Omega_e^2} |E_y|^2
      - i \frac{1}{2} \frac{k_\perp v_\perp^2 v_\|}{|\Omega_e|}
      \left( E_y^*E_z - E_y E_z^* \right)
      + v_\|^2 |E_z|^2 \right]
\end{multline}
where the first term in the bracket is transit-time damping (from
$\mu\grad B$ mirror force), the last term is Landau damping due to
parallel electric field, and the middle term is the combination of
these terms.  Next we stipulate that the suprathermal electrons (with
a ratio of at most $n_{sth}/n~10^{-5}$) do not significantly alter the
fields as found using the cold plasma approximation and the limits
used to derive the dispersion relation of Eq. ().  The fields may then
be related as,
\begin{equation}
  \label{eq:apb-3}
  E_z=\frac{\kpar\kperp{}}{1+\kperp{}^2} E_x
  \qquad
  E_x = \frac{-i \bar{k}^2}{\bar{\omega}} E_y
  \qquad
  E_y = \frac{\omega}{k_\perp c} B_z
\end{equation}
where $\bar{\omega}=\omega/|\Omega_e|$ and $\bar{k}=kc/\omega_{pe}$.
The power dissipated by the suprathermal electrons is then entirely
expressed in terms of the parallel component of the magnetic field,
\begin{multline}
  \label{eq:apb-4}
  P = -\omega \frac{|B_z|^2}{8 n}
  \int_0^{\infty} 2\pi v_\perp {\rm d}v_\perp
  \int_{-\infty}^{\infty} {\rm d}v_{\|} \delta\left(v_{\|} - \frac{\omega}{k_{\|}}\right)\ppx{f_0}{v_\|}
  \\
  \left[
    \frac{1}{4}\left(\frac{v_\perp^4}{V_{Ae}^2}\right)
    - 
    \frac{\kpar}{1+\kperp{}^2} \frac{\bar{k}^2}{\bar{\omega}}
    \left(\frac{v_\perp^2 v_\|}{V_{Ae}}\right)
    +
    \frac{\kpar^2\bar{k}^4}{\left(1+\kperp{}^2\right)^2\bar{\omega}^2}\left(v_\|\right)^2
    \right]
\end{multline}
where $V_{Ae}=c |\Omega_e|/\omega_{pe}\simeq 5\times 10^9$cm/s.  For
Maxwellian suprathermal distributions these integrals can easily be
done, yielding,
\begin{multline}
  \label{eq:apb-6}
  P = \frac{n_{sth}}{n}\frac{\omega|B_z|^2}{4\pi^{1/2}}\frac{\bar{\omega}}{\kpar}
  \frac{V_{Ae}^3}{v_{sth}^3}\frac{\bar{k}^4}{\left(1+\bar{k}_\perp^2\right)^2}
  \exp\left(-\frac{\bar{\omega}^2}{\kpar^2}\frac{V_{Ae}^2}{v_{sth}^2}\right)
  \\
  \times
 \left\{
 \left[1-\frac{v_{sth}^2}{V_{Ae}^2}\frac{(1+\kperp{}^2)}{2\bar{k}^2}\right]^2
 +\frac{v_{sth}^4}{V_{Ae}^4}\frac{(1+\kperp{}^2)^2}{4\bar{k}^4}
 \right\}
\end{multline}
where typical values of the ratio $V_{Ae}/v_{sth}\sim 3-10$.  By
examination of Eq. (\ref{eq:apb-6}) the transit time damping is
typically a little smaller than Landau damping, and interestingly acts
in a way as to reduce the damping rate.  Bell \textit{et
  al.}\cite{bell02} have published an analytical distribution function
given by,
\begin{equation}
  \label{eq:apb-bellf0}
  f_0^{Bell} = \frac{a}{v^4} - \frac{b}{v^5} + \frac{c}{v^6}
\end{equation}
where $a$, $b$, and $c$ are constants.  The integrals in
Eq. (\ref{eq:apb-5}) can also be calculated with
Eq. (\ref{eq:apb-bellf0}), assuming the functions go to zero more
quickly than $1/v^4$ outside of the range where data has been
collected.  We have computed these integrals and present the numerical
values of the damping rates resulting from these calculations below in
table \ref{tab:apb-1}.

For the Landau and transit time damping to be significant the phase
velocity of the waves must coincide with the velocity of the
suprathermal electrons.  This limits wave-vectors to specific ranges
to consider.  For example, whistler waves (with
$(m_e/m_i)^{1/2}<\bar{k}_{\|}<1$ and $\bar{k}_{\perp}<1$) have phase
velocities that match electrons with energies between 100 and 1500 ev.
In this case, the dispersion relation simplifies to
$\bar{\omega}^2=\kpar^2\bar{k}^2$.  Then to get the damping rate
$\gamma=-P/W$ we need the energy density of the waves,
\begin{equation}
  \label{eq:apb-5}
  W = \frac{1}{16\pi}\left[ \mvec{B}^*\cdot\mvec{B} + 
    \mvec{E}^*\ppx{}{\omega}\mvec{\epsilon}_h\mvec{E}\right]
\end{equation}
which for whistler waves is dominated by the magnetic field energy,
becoming simply $W\simeq(|B_\perp|^2+|B_z|^2/(16\pi)\simeq(1+\bar{k}_\|^2/\bar{k}_\perp^2)|B_z|^2/(8\pi)$.  Then the damping rate of whistler
waves with $(m_e/m_i)^{1/2}<\bar{k}_\|\sim\bar{k}_\perp<1$due to Maxwellian suprathermal electrons becomes,
\begin{equation}
  \label{eq:apb-7}
  \gamma = -\sqrt{\pi}\omega\frac{n_{sth}}{n}
  \frac{V_{Ae}^3}{v_{sth}^3}\bar{k}^5
  \exp\left(-\bar{k}^2\frac{V_{Ae}^2}{v_{sth}^2}\right)
\end{equation}
where the terms proportional to $v_{sth}^2/(2 \bar{k}^2 V_{Ae}^2)<1$
have been dropped.  On the other hand, quasi-electrostatic
lower-hybrid waves ($\bar{k}_{\perp}<1$ and
$\bar{k}_{\|}<\bar{k}_{\perp}(m_e/m_i)^{1/2}$) have phase velocities
that match the velocity of electrons with 10's of KeV, for which there
aren't many suprathermal electrons.  Thus, toward the magnetospheric
edge of the cavity, where such waves are prevalent, the collisional
damping will always dominate.

Finally, we quantitatively compare the collisionless damping rates to
the collisional damping rate for different values of
$(\kpar,\kperp{})$ in Table \ref{tab:apb-1}.  In this table we
calculate the collisionless damping rate in two ways.
$\gamma_{L-Bell}$ is calculated from Eq. (\ref{eq:apb-4}), where we
have computed the velocity space integrals using Bell's published
analytical distribution function given by Eq. (\ref{eq:apb-bellf0}),
and dividing by Eq. (\ref{eq:apb-5}) where we have used the cold
plasma expressions for $\epsilon_h$ and the fields in
Eq. (\ref{eq:apb-3}).  $\gamma_{L-Maxwellian}$ is calculated from
Eq. (\ref{eq:apb-6}) and Eq. (\ref{eq:apb-5}).  For whistlers this
rate becomes Eq. (\ref{eq:apb-7}).  For a suprathermal density we took
$n_{sth}=0.013$ and $v_{sth}$ corresponding to 150 eV.  For each of
these calculations we choose typical L=2 parameters of $B=0.04$ G and
$n=5\times 10^3$/cm$^3$.  The first case $(\kpar,\kperp{})=(1/7,1/7)$
corresponds to whistlers inside the cavity, where we see that the
collisionless damping is at least one order of magnitude smaller than
the collisional damping we have considered.  The second case
corresponds to a magnetosonic wave, and the third and the fourth cases
correspond to a lower-hybrid like wave characteristic of the
magnetospheric edge of the cavity.  In these last three cases the
damping rate due to suprathermal electrons is at least an order of
magnitude smaller than the collisional damping we have considered.

\begin{center}
\begin{table*}[ht]
  \hfill{}
  \begin{tabular}{|c|c|c|c|c|c|}
    \hline
    $\kpar$ & $\kperp{}$ & $\gamma_{Coll}$ (rad/s) & $E_{Res}$ (KeV) & $\gamma_{L-Bell}$ (rad/s)& $\gamma_{L-(Maxwellian)}$ (rad/s)
    \\ \hline\hline
    1/7 & 1/7 & 3$\times 10^{-2}$ & 0.3 & 5$\times 10^{-5}$ & 8$\times 10^{-4}$
    \\ \hline
    0.02 & 0.1 & 4$\times 10^{-3}$ & 0.2  & 1$\times 10^{-5}$ & 3$\times 10^{-5}$
    \\
    \hline
    1/7 & 1/2 & 1$\times 10^{-1}$ & 1.4 & 2$\times 10^{-3}$ & 2$\times 10^{-4}$
    \\
    \hline
    1/7 & 1.5 & 0.3 & 1.9 & 2$\times 10^{-2}$ & 5$\times 10^{-5}$
    \\ \hline
  \end{tabular}
  \hfill{}
  \caption{Damping rates assuming $B=0.04$ G and $n=5\times 10^3$/cm$^3$.  For $\gamma_{L-(Maxwellian)}$ we take $n_{sth}=0.013$/cc and $v_{sth}$ corresponding to 150 eV.  }
  \label{tab:apb-1}
\end{table*}
\end{center}

\section{Estimation of NL Scattering Rate}
\label{sec:app}

In this appendix we simplify the NL induced scattering rate of the
main text.  We start with a general form of the NL induced scattering
rate as published in \citet{ganguli10},
\begin{multline}
  \gamma_{NL} = \frac{1}{\omega_{k2}} \frac{\bar{k}_2^2}{1+\bar{k}_2^2}
  \sum_{k_1} \frac{|E_{k1}|^2}{4\pi n m_e} 
  \frac{\left| \mvec{k}_1 \times \mvec{k}_2\right|^2_{\|}}{k_{\perp 1}^2 k_{\perp 2}^2}
  \\
  \times
  \frac{(\mvec{k}_2 - \mvec{k}_1)^2 \bar{k}_1^2}{1+\bar{k}_1^2} 
  \frac{{\rm Im} \epsilon^{e}_{k_1-k_2} |\epsilon^i_{k_1-k_2}|^2}{
    \left|\epsilon^e_{k_1-k_2} + \epsilon^i_{k_1-k_2}\right|^2}
\end{multline}
where, assuming a fluid model for the ions and drift kinetic model for the electrons,
\begin{equation}
  \begin{split}
    \epsilon^{i}_{k_1-k_2} &= \sum_s \frac{\omega_{ps}^2}{\Omega_s^2 - (\Delta \omega)^2}
    \\
    \epsilon^{e}_{k_1-k_2} &= \frac{\omega_{pe}^2}{(\Delta k)^2 v_{te}^2/2} \left(1 +  \zeta Z(\zeta)\right)
  \end{split}
\end{equation}
and $\epsilon^i_{k_1-k_2}$ is a real function while
$\epsilon^e_{k_1-k_2}$ has an imaginary part from the dispersion
function.  Here $\Delta \omega = \omega_{k_1} - \omega_{k_2}$, $\Delta
k=|\mvec{k}_{1}-\mvec{k}_2|$ and $\Delta k_z=k_{1z} - k_{2z}$ are the
frequency and wave vectors of the beat wave, and the subscripts 1 and
2 denote the mother and daughter waves respectively.  For NL wave
scattering by magnetized electrons the condition
\begin{equation}
  \label{eq:a3}
  \zeta = \frac{(\omega_{k1}-\omega_{k2})}{|k_{2z}-k_{1z}|}\lesssim 1
\end{equation}
must be satisfied, \textit{i.e.} there must be enough particles that
resonate with the beat wave structure and $\zeta Z(\zeta)\sim 1$.  To
maximize the NL rate the inequality
$|\epsilon^{i}_{k_1-k_2}|>|\epsilon^{e}_{k_1-k_2}|$ is necessary.
This inequality is satisfied under two cases:
\begin{equation}
  \label{eq:a4}
  (1)\,\Omega_s<|\Delta\omega|<\Delta kC_s \rightarrow (\Delta k)^2\rho_s^2 >1
\end{equation}
is the subsonic condition (meaning the beat wave has a frequency below
the ion-sound wave frequency), and 
\begin{equation}
  \label{eq:a5}
  (2)\,|\Delta\omega-\Omega_i| = (\Delta k)^2\rho_s^2,\qquad (\Delta k)^2\rho_s^2 < 1
\end{equation}
is the ion cyclotron condition (meaning the beat wave has a frequency
below the electrostatic ion cyclotron wave frequency), where $C_s^2 = 2T_e/m_i$ and $\rho_s =
C_s/\Omega_i$.  When (\ref{eq:a3}) and one of the conditions
(\ref{eq:a4}) or (\ref{eq:a5}) is satisfied the scattering becomes,
\begin{multline}
  \gamma_{NL} \sim \frac{\Omega_e^2}{\omega_{k2}} \frac{\bar{k}_2^2}{1+\bar{k}_2^2}
  \sum_{k_1} \frac{\omega_{pe}^2}{\Omega_e^2} \frac{|E_{k1}|^2}{4\pi n T_e} 
  \frac{\left| \mvec{k}_1 \times \mvec{k}_2\right|^2_{\|}}{k_{\perp 1}^2 k_{\perp 2}^2}
  \frac{\bar{k}_1^2}{1+\bar{k}_1^2}
\end{multline}
Considering a broadband spectra of width $\delta\omega\gg\Delta\omega$
we simplify the sum over all $\mvec{k}$ by considering only the waves
contributing to the summation given the conditions of
Eq. (\ref{eq:a3}) and (\ref{eq:a4}) or (\ref{eq:a5}).  Since the NL
scattering rate is maximum for large angle scatterings we estimate
that $|\mvec{k}_1\times\mvec{k}_2|^2_{\|}/(k_{\perp 1}^2k_{\perp
  2}^2)\sim 1$.  For the subsonic case (case 1 above) the NL
scattering rate may be estimated by the fraction of interacting waves
($\Delta\omega/\delta\omega\simeq\Delta k C_s/\delta\omega$) under the
condition of Eq. (\ref{eq:a3}) and (\ref{eq:a4}),
\begin{equation}
  \label{eq:a9}
  \gamma_{NL} \sim \frac{\Omega_e^2}{\omega_{k2}} \frac{\bar{k}_2^2}{1+\bar{k}_2^2}
  \frac{W_1}{n_0 T_e}
  \frac{\eavg{\bar{k}_1}^2}{1+\eavg{\bar{k}_1}^2}
  \frac{\omega_{LH}}{\delta\omega}
  \Delta \bar{k} \beta_e^{1/2}
\end{equation}
where,
\begin{equation}
  \label{eq:a8}
  \frac{W_1}{n T_e} = \frac{\omega_{pe}^2}{\Omega_e^2} \sum_{k_1} \frac{|E_{k1}|^2}{8\pi n T_e} 
  \simeq \frac{B_1^2}{8\pi n T_e} \eavg{\bar{k}_1}^2
\end{equation}
and $\eavg{\bar{k}_1}$ is the average value of $\bar{k}_1$.  Note that
$N\omega=W_1$ and Eq. (\ref{eq:a8}) gives the relation between the
plasmon number density and the amplitude of the fluctuating magnetic
field.

For the ion cyclotron case (case 2 above) the beat wave frequency is
the ion-cyclotron frequency, but the $\mvec{k}_1$'s that contribute to
the summation are restricted by the condition (\ref{eq:a5}), $\Delta
k^2\rho_s^2<1$, so that we can write the NL scattering rate as
\begin{equation}
  \label{eq:a10}
  \gamma_{NL}\sim
  \frac{\Omega_e^2}{\omega_{k2}} \frac{\bar{k}_2^2}{1+\bar{k}_2^2}
  \frac{W_1}{n_0 T_e}
  \frac{\eavg{\bar{k}_1}^2}{1+\eavg{\bar{k}_1}^2}
  \frac{\omega_{LH}}{\delta\omega}
  \Delta\bar{k}^2\beta_e\left(\frac{m_i}{m_e}\right)^{1/2}
\end{equation}
Now, we evaluate these conditions near the lower-hybrid resonant
surface in the magnetosphere to determine whether the rate for
scattering of lower-hybrid waves to magnetosonic/whistler waves is
governed by Eq. (\ref{eq:a9}) or (\ref{eq:a10}).  Here the plasma is
almost all hydrogen and $\beta_e\sim6\times10^{-5}$ so that $(\Delta
k)^2\rho_{sH}^2 = (\Delta \bar{k})^2 \beta_e m_H/m_e\sim
0.1(\Delta\bar{k})^2$.  In this case, to meet the condition of
Eq. (\ref{eq:a4}) would require $(\Delta\bar{k})^2>10$, which from
ray-tracing (as can be seen in Figure \ref{fig:kperp}) cannot be
satisfied .  However it is possible to meet the condition
(\ref{eq:a5}) since $(\Delta\bar{k})^2<10$, and the condition
(Eq. \ref{eq:a3})
\begin{equation}
  \label{eq:a11}
  \frac{\Delta\omega}{|k_{2z}-k_{1z}|v_{te}} = \frac{\Omega_H}{\Delta k_z v_{te}} \le 1
\end{equation}
which evaluates to $(\Delta\bar{k}_z)^2 \ge 10 m_e/m_H$, is not too
restrictive.  Therefore Eq. (\ref{eq:a10}) is the proper estimate of
the NL scattering rate for the formation of the magnetospheric cavity,
which is somewhat smaller than the rate as would be given by
Eq. (\ref{eq:a9}).  The rate given by Eq. (\ref{eq:a9}) is applicable
to the scattering of lower hybrid waves to magnetosonic/whistler waves
in the ionosphere where oxygen plasma dominates and the main parameter
$\beta_e m_o/m_e$ is of the order of unity.

To arrive at Eq. (\ref{eq:agnl}) in the text we consider the
large-angle NL scattering from lower-hybrid waves to whistler waves so
that $\Delta\bar{k}\sim\eavg{\bar{k}}$ in Eq. (\ref{eq:a10}).  From
ray-tracing waves with $\omega\sim\omega_{LH}$ we find that $k_{\perp
  1}>k_{\|1}$ and we estimate the scattered wave vector $k_{\perp
  2}\sim\eavg{k_{\perp 1}}$.  With these assumptions and
Eq. (\ref{eq:a8}) we arrive at
\begin{equation}
  \label{eq:apagnl}
  \gamma_{NL} \sim \omega_{LH} \left(\frac{M}{m}\right)^{3/2}  \frac{\omega}{\delta \omega} 
  \frac{\eavg{\kperp{1}}^8}{\left(1+\eavg{\kperp{1}}^2\right)^2} \frac{B^2_{1}}{B_0^2}
\end{equation}
where the subscript 1 has been dropped in Eq. (\ref{eq:agnl}).

\begin{acknowledgments}
This work is supported by the Naval Research Laboratory Base Program.
\end{acknowledgments}

%

\end{document}